\documentclass[letterpaper,10pt]{article} 

\usepackage{osameet3} 

\usepackage{amsmath,amssymb}
\usepackage[colorlinks=true,bookmarks=false,citecolor=blue,urlcolor=blue]{hyperref} 

\begin{document}

\title{Optical Adaptive LMS Equalizer with an Opto-electronic Feedback Loop}

\author{Xumeng Liu, Xingyuan Xu and Arthur James Lowery}
\address{Electro-Photonics Laboratory, Dept. Electrical and Comp. Systems Eng., Monash University, Clayton, VIC 3800, Australia}
\email{xumeng.liu@monash.edu}

\copyrightyear{2021}

\begin{abstract}
We propose a photonic adaptive Least-Mean-Squares equalizer with an opto-electronic feedback loop to determine the updating of an optical Finite-Impulse-Response filter's weights, enabling dispersion compensation introduced by 30-km fiber based on our photonic integrated chip. 
\end{abstract}

\section{Introduction}
Optical equalization of optical communications systems has been used since the 1990s; for example, adding dispersion-compensating modules (DCMs) that contain dispersion compensating fibers (DCFs) \cite{gruner2005dispersion}, fiber Bragg gratings (FBGs) \cite{eggleton2000integrated} or Mach-Zehnder interferometers (MZIs) \cite{bohn2002tunable,doerr2004simple,mikhailov2006fiber}. However, over the last decade, Digital Signal Processing (DSP) has become the main technique to improve the performance of long-haul optical fiber communications systems \cite{Savory2007edc}. In many systems designs, DSP has replaced designed-in dispersion compensation that used DCMs throughout the line; this has the advantage that it replaces the link-specific `backroom' engineering design, with `plug-and-play' adaptive equalization. Being adaptive, it can also support optically-switched networks. 

Unfortunately, DSP chipsets consume significant power and are also based on channel-wise processing \cite{li2008electronic}. This is in contrast to the optical DCMs that they replace, which can `process' 10-100's of wavelength channels simultaneously in the optical domain, and only require a single optical amplifier, consuming a few watts of electrical power. Adaptive optical equalization could give the best of both worlds: a large processing bandwidth together with `plug-and-play' operation. 

A traditional equalizer for wireless systems is the Least-Mean Squares Finite-Impulse Response (LMS-FIR) equalizer. In earlier work, we have shown that the LMS-FIR equalizer, implemented in digital signal processing, is able to select and optimize a channel in a multi-subcarrier system \cite{du2011no}. Our aim in this work is to transfer the majority of the DSP into the optical domain, so we can potentially reduce power consumption and increase the processing bandwidth. Our FIR PIC could form the basis of this optical equalizer \cite{xie2018picosecond}.

In this paper, we develop an adaptive optical LMS equalizer based on our previously reported photonic integrated circuit (PIC) FIR filter \cite{xie2018picosecond}. Using simulations, we develop methods to implement the feedback loop using electro-optic techniques, with the mixing properties of a high-speed square-law photodetector to derive signals that can be used to update the tap weights of the filter. We demonstrate adaptive equalization of chromatic dispersion (CD) affecting Quadrature Phase Shift Keying (QPSK) signal, which shows that the equalizer converges to optimal weights to compensate CD. Although we show one channel being equalized, the FIR filter could equalize dispersion over multiple channels if they have similar dispersions.

\section{System Design}

\begin{figure}[htbp]
  \centering
  \includegraphics[width=\textwidth]{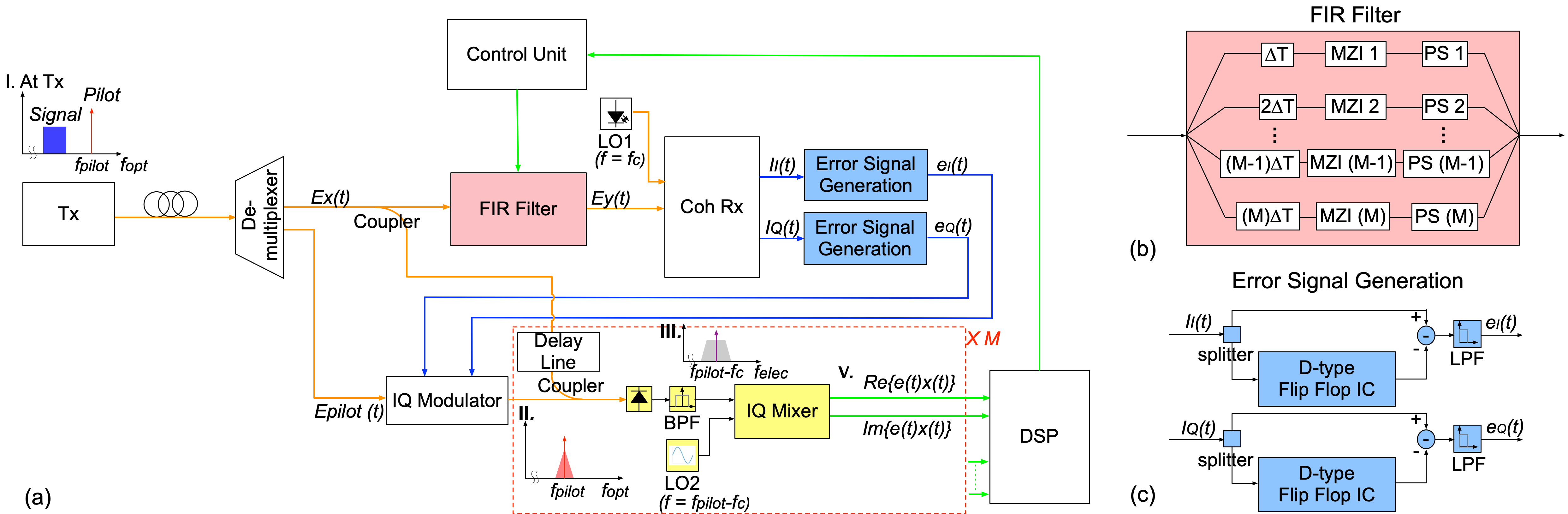}
\caption{(a) Schematic of the adaptive optical LMS equalizer system (in decision-directed mode). Magnified views of: (b) functional equivalent of the FIR filter, (c) error signal generation circuits.}
\label{schematic}
\end{figure}

The adaptive optical LMS-FIR equalizer system is illustrated in Fig. \ref{schematic}. A data-carrying signal is transmitted along with a pilot tone; after de-multiplexing, the signal is processed by an optical filter then a coherent receiver, and the pilot tone serves as an optical carrier for the feedback processing. The optical FIR filter is based on a PIC that we have fabricated and tested \cite{xie2018picosecond}. It uses separate delays with MZIs and optical phase shifters on each arm, to allow its complex weights to be tuned. The equalization is first set in a training mode, then tracks changes in a decision-directed mode, in which the symbol error rate is low enough to generate an error signal by subtracting the estimated signal from the actual signal, e.g. by using a high-speed D-type flip-flop for two-level signal, then subtracting this digital value from the direct analog output of the equalizer, as shown in Fig. \ref{schematic}(c). This is done for both the inphase output and quadrature output of the coherent receiver. The error signal is modulated onto the pilot tone by an IQ modulator, then added to a delayed version of the input signal and passed to a photodiode. 
The photodiode multiplies the error by the delayed signals, using its square-law characteristic. The unwanted DC mixing components are removed by a passband filter, and the wanted passband signal (\(e(t)x(t)\)) is downconverted using a microwave IQ mixer and then filtered out with a low-pass filter. These signals could be sent to the pilot's modulator directly; however, it is more flexible to implement this part of the feedback loop using DSP for the integration and adjustment of the feedback factor (\(w_{i+1} = w_i + {\mu}e(t)x(t)\)).

\section{Results and Discussion}
To verify the effectiveness of the adaptive optical LMS equalizer system, we simulated the equalization of a 40-GBd NRZ QPSK signal. The FIR equalizer has 15 taps with the inter-tap delay of 6.25 ps, which is consistent with our chip. Here, it acts as a fractionally-spaced equalizer with an FSR of 160 GHz. 
To illustrate the compensation of CD, we simulated the transmission channel as a single-mode standard fiber (D = 16 ps/nm/km) with a length of 25 km. 



\begin{figure}[htbp]
\centering
\includegraphics[width=\textwidth, height=2.7in]{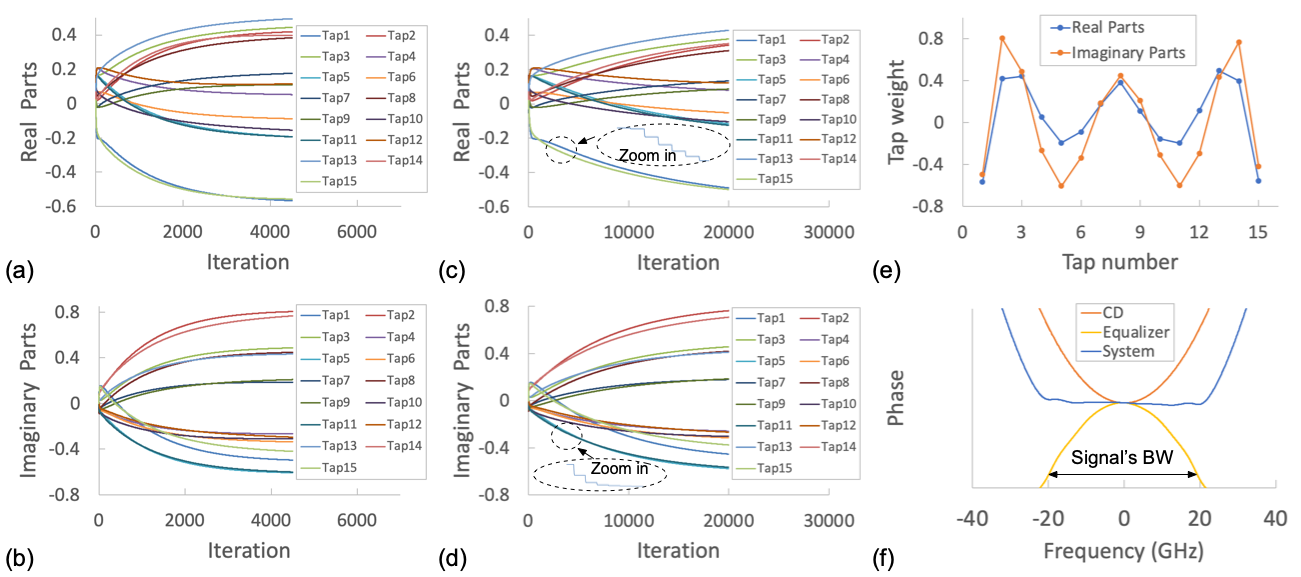}
\caption{Tap weights (Real and Imaginary parts) against accumulated simulation runs (iterations): (a-b) all taps updated each run and (c-d) taps updated sequentially, with 10 runs per tap before moving to the next tap. (e) Tap weights and (f) phase response when converged. }
\label{weight_CD}
\end{figure}

Fig. \ref{weight_CD}(a-d) shows the weights of the taps as they converge, using two different approaches. The feedback gain has been selected to give stability with convergence to a steady state. In Fig. \ref{weight_CD}(a) and (b), the tap weights of the 15 taps are updated after every 256 symbols, using 15 parallel error calculation circuits. Obviously, this would be expensive in hardware; however, it is possible for the weights to be sequentially updated, one by one, with each update being the duration of 10 simulation runs with 256 symbols in each run. Fig. \ref{weight_CD}(c) and (d) shows that the tap weights still converge to the optimum values, although convergence requires more iterations. This sequential updating method requires only one feedback loop for the multi-tap optical equalizers, which would lead to a reduction in hardware. We note that the increased convergence time of the latter approach is dependent on the number of taps and simulation runs in each turn.
Fig. \ref{weight_CD}(e) and (f) shows tap weights and the corresponding phase responses when converged. The fiber has a parabolic phase response (grey); we see that within the signal-spectrum's bandwidth (40 GHz), the adaptive equalizer converges to a parabolic response of the opposite sense, to give a near-flat phase response of the system (blue).

\begin{figure}[htbp]
\centering
\includegraphics[width=\textwidth]{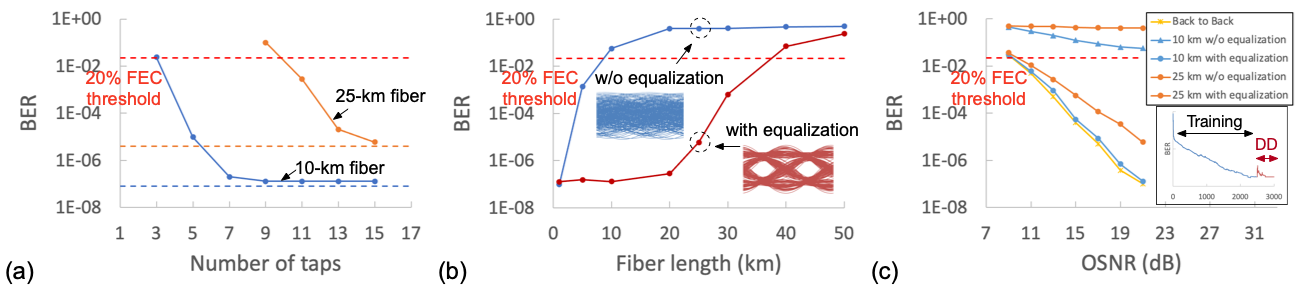}
\caption{Performance of CD compensation: (a) BER versus number of taps, (b) BER versus fiber length, (c) BER versus OSNR. }
\label{OSNR}
\end{figure}

To test the performance of the equalizer system, we have introduced CD by different lengths of standard single-mode fiber and added noise to the system. Fig. \ref{OSNR}(a) shows the influence of the number of taps on CD compensation when OSNR = 21 dB. For 10-km fiber, a 3-tap equalizer can compensate dispersion-introduced intersymbol interference; however, at least 11 taps are required for 25-km fiber to achieve BER below 20\% FEC threshold, with a delay difference of 62.5 ps over all of the taps. For comparison, the differential group delay of 25 km of fiber is 128 ps, which is equivalent to 22 taps.

Fig. \ref{OSNR}(b) shows the BER performance without and with equalization for 1 to 50-km of standard single-mode fiber when implementing 15 taps at OSNR = 21 dB. Without equalization, the effect of CD is severe for lengths of more than 10 km; however, the equalizer system achieves good CD compensation performance with the aid of training, and can compensates 20 km of fiber without a penalty, and up to 30 km at the FEC limit. 

Fig. \ref{OSNR}(c) shows the BER versus OSNR of the signals at the 15-tap equalizer output for back-to-back, 10-km fiber and 25-km fiber. The BER is below the 20\% FEC threshold when the OSNR is \(>\)11 dB. This indicates that this adaptive equalizer system has good tolerance to noise. The BER convergence for 25-km fiber when OSNR = 21 dB is also shown. The BER decreases to \(10^{-7}\) during training (2000 iterations), after which the tap weights have converged. We found that when switched to decision-directed mode (see inset), there is a small temporary increase in error rate, before returning to the same low level as during training. This glitch is because the data is derived from decisions on the received signals, which are slightly misaligned to the training data.

\section{Conclusions}
In this paper, we have proposed and simulated a novel adaptive optical LMS equalizer with a feedback loop using electro-optic techniques. Simulations show that this equalizer is capable of compensating CD with the feedback control signal partially calculated using the mixing properties of photodiodes. In contrast to conventional digital LMS equalizers, this adaptive optical LMS equalizer realizes part of signal processing using a photonic FIR filter, which could consume very little energy in the steady state, if controlled by electrostatic techniques \cite{boes2018status}. We also show that a single feedback loop can be time-multiplexed between the calculations for the multiple weights of the FIR filter, to save a large amount of hardware. Although we have shown one channel being equalized, the FIR filter could equalize multiple channels if they have similar dispersions.

\section*{Acknowledgements}
This work was supported by the Australian Research Council's Discovery Project scheme (DP190101576). We should like to thank VPIphotonics.com for the use of VPItransmissionMaker.


\bibliographystyle{osajnl}
\bibliography{Reference}

\end{document}